\documentclass[11pt]{article}
\usepackage{amssymb, latexsym, amsmath}
\usepackage{amsfonts, graphics}
\def\R{\mathbb R}

\def\be{\begin{equation}}
\def\ee{\end{equation}}
\def\bea{\begin{eqnarray}}
\def\eea{\end{eqnarray}}
\def\beas{\begin{eqnarray*}}
\def\eeas{\end{eqnarray*}}

\def\open#1{\setbox0=\hbox{$#1$}
\baselineskip = 0pt
\vbox{\hbox{\hspace*{0.4 \wd0}\tiny $\circ$}\hbox{$#1$}}
\baselineskip = 11pt\!}

\newcommand{\prfe}{\hspace*{\fill} $\Box$

\smallskip \noindent}

\begin{document}
\sloppy
\newtheorem{theorem}{Theorem}[section]
\newtheorem{definition}[theorem]{Definition}
\newtheorem{proposition}[theorem]{Proposition}
\newtheorem{example}[theorem]{Example}
\newtheorem{remark}[theorem]{Remark}
\newtheorem{cor}[theorem]{Corollary}
\newtheorem{lemma}[theorem]{Lemma}

\renewcommand{\theequation}{\arabic{section}.\arabic{equation}}

\title{Galactic dynamics in MOND---Existence of equilibria
       with finite mass and compact support}

\author{Gerhard Rein\\
        Fakult\"at f\"ur Mathematik, Physik und Informatik\\
        Universit\"at Bayreuth\\
        D-95440 Bayreuth, Germany\\
        email: gerhard.rein@uni-bayreuth.de} 

\maketitle
\begin{abstract}
We consider a self-gravitating collisionless gas
where the gravitational interaction is modeled according
to MOND (modified Newtonian dynamics). For the resulting
modified Vlasov-Poisson system we establish
the existence of spherically symmetric
equilibria with compact support and finite mass.
In the standard situation where gravity is modeled by
Newton's law the latter properties
only hold under suitable restrictions on the prescribed
microscopic equation of state. Under the MOND regime
no such restrictions are needed.
\end{abstract}
\section{Introduction}
\setcounter{equation}{0}
One of the intriguing mysteries in current astrophysics is 
the possible existence and nature of dark matter around galaxies.
Observed rotational velocities of stars and gas in typical
spiral galaxies seem to be larger than
the ones predicted from the gravitational potential
of the directly observable  matter. 
This is one manifestation of the so-called
{\em missing mass problem}. 
The resolution of this problem
which is currently favored by a majority
in the astrophysics community is that a galaxy is typically
surrounded by a spherical halo of dark matter which provides
the missing mass. This dark component
is supposed to outweigh the visible matter by a factor
of the order 10.
There are other indirect arguments for the
existence of dark matter, but so far there is no direct observational
evidence for its existence nor a consistent physical theory
which predicts its existence.

Some 30 years ago {\sc M.~Milgrom} proposed MOND
(modified Newtonian dynamics) which
predicts the observed rotational velocities in galaxies from
the visible matter without invoking a dark component. 
The basic MOND paradigm can be expressed as follows.
If a particle (for example a star) 
would in Newtonian gravity experience an acceleration $g_N$,
then according to MOND  the particle experiences an
acceleration $g$ which obeys the relation
\be \label{Mpara}
\mu(|g|/a_0) g = g_N.
\ee
Here $a_0\approx 10^{-10} m/s^2$
is a new physical constant, 
an acceleration far below typical accelerations
in the solar system, and $\mu$ is an interpolating
function such that
\be \label{mucond1}
\mu(\tau) \to 1\ \mbox{for}\ \tau >>1\ \mbox{and}\
\mu(\tau) \to \tau\ \mbox{for}\ \tau << 1.
\ee
If $|g|<<a_0$ then MOND predicts an
acceleration $|g| \sim \sqrt{|g_N|}$ which is much larger than
the Newtonian prediction, while the two coincide for $|g|>>a_0$. 
It is a surprising empirical fact 
that a modification of Newtonian mechanics with
one free parameter, namely $a_0$, seems to 
correctly predict the rotation curves for a large variety
of galaxies. We refer to \cite{milgr1,milgr2} for very readable
introductions to MOND,
and to \cite{mcgaugh} for an in-depth discussion of MOND,
where the observational support for dark matter
and a very large part of the corresponding literature 
are reviewed. Background on dark matter is also found in \cite{BT}.

In the present paper we investigate self-consistent
mathematical models for galaxies or globular clusters
where the gravitational interaction is to obey the MOND
paradigm. A galaxy is often modeled as a 
large ensemble of particles which interact only
via gravity.
This results in the Vlasov or Collisionless Boltzmann equation,
coupled to a suitable field equation for gravity.
If $-\nabla_x U$  denotes the gravitational field then the Vlasov equation
reads
\be \label{vlasov}
\partial_t f + v \cdot \nabla_x f - \nabla_x U \cdot \nabla _v f=0.
\ee
Here $f=f(t,x,v)\geq 0$ is the density of the particle ensemble
in phase space and $t\geq 0$, $x,v\in \R^3$ stand for time, position,
and velocity. The density in phase space induces the spatial
mass density
\be \label{rhodef}
\rho (t,x) = \int f(t,x,v)\,dv.
\ee
We close the system with the
modified Poisson equation
\be \label{Mpoisson}
\mathrm{div} \left[\mu(|\nabla U|/a_0) \nabla U \right] = 4 \pi \rho
\ee
which  determines the gravitational potential $U$ 
in terms of $\rho$.
If $\mu$ satisfies (\ref{mucond1}) this field equation implies
the MOND paradigm (\ref{Mpara}), cf.\ \cite[Eqn.(17)]{mcgaugh};
this is just one possible implementation of the
MOND paradigm in terms of a consistent physical theory.
The field equation (\ref{Mpoisson}) is non-linear while 
in the Newtonian situation the non-linearity in the resulting
Vlasov-Poisson system arises through the coupling with the Vlasov
equation.
Astrophysical background on the Vlasov-Poisson system
can be found in \cite{BT}. 
We refer to the system (\ref{vlasov}), (\ref{rhodef}), (\ref{Mpoisson})
as the MONDian Vlasov-Poisson system. In passing we note that
the usual normalizing condition
\be \label{bc}
\lim_{|x|\to \infty} U(t,x) =0
\ee
does not work in the MOND case; we will 
come back to the issue of the behavior at infinity.

In the present paper we prove the existence of
spherically symmetric steady states
of this system which have finite mass and compact spatial support.
In this context it is instructive to generalize the
class of interpolating functions $\mu$ as follows. We require
that $\mu(\tau) \to  1,\ \tau >> 1$ and
$\mu(\tau) \to \tau^\alpha,\ \tau << 1$ for some $\alpha\in[0,1]$.
Genuine MOND corresponds to the choice $\alpha =1$,
while the choice $\alpha =0$ or more precisely $\mu=1$
yields the standard Vlasov-Poisson system
of Newtonian galactic dynamics. 
The MOND case $\alpha =1$ and a case with $\alpha = 1-\epsilon$
and $\epsilon >0$ very small should 
be hard to distinguish observationally,
but mathematically these two cases behave quite differently.

The approach we use is known from the Newtonian case. 
The system under investigation is---by a suitable ansatz---reduced 
to the modified Poisson equation (\ref{Mpoisson}) where the
right hand side becomes a function of the unknown potential
$U$. The latter functional relation is determined by
the assumed microscopic equation of state, i.e., by the functional 
dependence of $f$ on the particle energy $E=\frac{1}{2}|v|^2 + U(x)$
and possibly other local conserved quantities.
In the Newtonian case the resulting steady state has the physically
required properties of finite mass and compact support only 
under suitable
restrictions on the equation of state.
We refer to \cite{RammRein} for a quite general, sufficient 
such condition. The main result of the present paper is the following.
For $\alpha \in [0,1[$ the same condition as in the 
Newtonian case guarantees finite mass and compact support,
while for the genuine MOND case $\alpha=1$ no such condition
(beyond technical assumptions) is needed.

The paper proceeds as follows. In the next section we 
make precise the basic set-up for constructing spherically symmetric
steady states of the MONDian Vlasov-Poisson system.
In particular, we establish Jeans' Theorem, which says that in 
the spherically symmetric case the particle distribution function must be a 
function of the particle energy and modulus of angular momentum,
we reduce the stationary system to a
non-linear equation for the potential $U$, and we 
prove a corresponding existence result.
In Section~\ref{sec_finite} we investigate the question of finite mass
and compact support of the steady states.  In a final section
we discuss the asymptotics at spatial infinity
and the question whether associated potential energies are finite.
We consider steady states
resulting from a Maxwellian ansatz.
In the Newtonian situation they
have infinite mass and extent, but
in the genuine MOND regime $\alpha=1$ their mass is finite. 
Finally, we extend our results to the MONDian Euler-Poisson 
system where matter is modeled as
an ideal, compressible fluid.

To conclude this introduction we recall that for the Newtonian
Vlasov-Poisson system there exists a satisfactory existence theory
for the initial value problem including a global-in-time
existence result for smooth initial data. Also the
stability properties of steady states are by now quite well understood,
and we refer to \cite{GR1,GR2,LMR,Rein07} and the references there.  
Whether these results persist under
the modification of the Poisson equation is an
interesting open problem.
The results of the present paper are mathematically fairly straightforward
which is in part due to the flexibility of the analysis in 
\cite{RammRein}, but our paper is 
intended only as a first step towards a better mathematical understanding
of MONDian modifications of stellar dynamics.

\section{Spherically symmetric steady states: The basic set-up}
\label{sec_setup}
\setcounter{equation}{0}

We first make precise our assumptions on the interpolating function
$\mu$ which appears in the basic MOND paradigm (\ref{Mpara}) and in the
modified Poisson equation (\ref{Mpoisson}).

\smallskip

\noindent
{\bf Assumptions on $\mu$.} Let $\mu \in C([0,\infty[) \cap C^1(]0,\infty[)$
be increasing with 
\be \label{muass}
\lim_{\tau \to \infty} \mu(\tau) =1,\qquad 
\lim_{\tau \to 0} \tau^{-\alpha} \mu(\tau) =1
\ee
for some $\alpha \in [0,1]$. 
Only the case $\alpha =1$ corresponds to the genuine
MOND theory. Furthermore we normalize the constant $a_0$ to unity, $a_0 =1$,
since its value does not affect the mathematical analysis.

We are interested in  steady states of the system 
(\ref{vlasov}), (\ref{rhodef}), (\ref{Mpoisson})
which are spherically symmetric, i.e.,
$f(x,v)=f(Ax,Av)$ for all orthonormal matrices $A$ and $x,v \in \R^3$.
A spherically symmetric distribution function depends only on the variables
\be \label{sphsymmvar}
r :=|x|,\ w := \frac{x \cdot v}{r},\  L := |x\times v|^2;
\ee
here $w$ is the radial velocity and $L$ the modulus of angular momentum squared.
The induced spatial density and
the potential are spherically symmetric in the sense that, by abuse of notation,
$\rho(x)=\rho(r)$ and $U(x)=U(r)$. Under these symmetry assumptions
the modified Poisson equation
(\ref{Mpoisson}) takes the form
\[
\frac{1}{r^2}\left(r^2 \mu(|U'|) U'\right)' = 4 \pi \rho.
\]
We integrate once to find that
\[
\mu(|U'|) U'  = \frac{m}{r^2},
\]
where 
\be \label{mdef}
m(r):= 4\pi \int_0^r s^2 \rho(s)\, ds
\ee
is the mass within the ball of radius $r$.
Since $\mu >0$ and $m \geq 0$, potentials solving the above equation
are increasing so that the modulus inside $\mu$ can be dropped. Now
we observe that the mapping $\tau \mapsto \tau \mu(\tau)$ is strictly
increasing on $[0,\infty[$ and onto $[0,\infty[$. Let $\zeta$ denote
its inverse, i.e.,
$\zeta : [0,\infty[ \to [0,\infty[$
is one-to-one, onto, strictly increasing,
$\zeta \in C([0,\infty[) \cap C^1(]0,\infty[)$, and
\be \label{zetaprop}
\zeta(\tau \mu(\tau))=\tau\ \mbox{for}\ \tau \geq 0,
\quad \lim_{\sigma \to 0} \sigma ^{-1/(1+\alpha)} \zeta(\sigma)=1,
\quad \lim_{\sigma \to \infty} \sigma ^{-1} \zeta(\sigma)=1.
\ee
The asymptotic properties of $\zeta$ follow from those of
$\mu$, cf.\ (\ref{muass}). 
Using this function we rewrite the spherically symmetric
modified Poisson equation in the form in which it will be 
solved, namely
\be \label{Mradpoisson}
U'(r) = \zeta\left(\frac{m(r)}{r^2}\right),\ r>0.
\ee
In the new variables adapted to the spherical symmetry (\ref{rhodef})
becomes
\be \label{rhodefss}
\rho(r) = \frac{\pi}{r^2} \int_{-\infty}^\infty\int_0^\infty f(r,w,L)\, dL\, dw.
\ee
The characteristic system 
\[
\dot x = v,\quad \dot v = -\nabla U(x)
\]
of the Vlasov equation (\ref{vlasov}) can be rewritten as
\be \label{charsrwL}
\dot r = w,\quad \dot w = \frac{L}{r^3} - U'(r),\quad \dot L = 0.
\ee
A spherically symmetric steady state of the MONDian Vlasov-Poisson
system is by definition a triple $(f,\rho,U)$
such that $f=f(r,w,L) \geq 0$ is measurable, 
$\rho$ satisfies (\ref{rhodefss}) and $r^2\rho$ is locally integrable 
on $[0,\infty[$, $U$ is differentiable
on $]0,\infty[$ and satisfies (\ref{Mradpoisson}),
and $f$ is constant along characteristics (\ref{charsrwL}) which must
exist uniquely for initial data $r>0$, $w\in \R$, and $L>0$.

Since the potential $U$ is time-independent the particle energy
\be \label{parten}
E=E(x,v) := \frac{1}{2} |v|^2 + U(x) = \frac{1}{2} w^2 + \frac{L}{2 r^2} + U(r)
\ee
is constant along characteristics, and due to spherical symmetry the
same is true for $L$.
Hence any function of the form
\be \label{ansatz}
f=\Phi(E,L)
\ee
with a suitable, prescribed function $\Phi\geq 0$ satisfies the Vlasov
equation in the sense of being constant along characteristics.
We show that by making the ansatz (\ref{ansatz}), no spherically
symmetric steady states are lost, a fact 
which in the Newtonian case is known as Jeans' Theorem, 
cf.\ \cite[Thm.~2.2]{BFH}. 
\begin{proposition} \label{jeans}
Let $(f,\rho,U)$ be a stationary, spherically symmetric solution of
(\ref{vlasov}), (\ref{rhodef}), (\ref{Mpoisson}).
Then $f$ is of the form (\ref{ansatz}).
\end{proposition}
{\bf Proof.} For $r>0$, any characteristic curve remains in a plane of constant
$L$ which we can take positive; $L$ is now kept fixed. 
In such a plane the effective characteristic system
can be rewritten as
\be \label{charseff}
\dot r = w,\quad \dot w = - \Psi_L'(r)
\ee
with effective potential 
\[
\Psi_L(r):= \frac{L}{2 r^2} + U(r),\ r>0,
\]
in particular, 
\[
E = E(r,w) = \frac{1}{2} w^2 + \Psi_L (r).
\]
The level sets of the function $E$ in the half plane $r>0, w\in \R$
are invariant under the flow of (\ref{charseff}), and since $f(\cdot,\cdot,L)$
is constant on the orbits of (\ref{charseff}) the assertion of the proposition
follows if each level set of $E$ consists of a single orbit.

In order to see the latter we analyze the effective potential $\Psi_L$.
We first show that $r^2 U(r) \to 0$ as $r\to 0$ so that
$\Psi_L(r) \to \infty$ as $r\to 0$. Indeed, for $0<r<1$,
\[
r^2 U(r) = r^2 U(1) - r^2 \int_r^1 \zeta\left(\frac{m(s)}{s^2}\right)\, ds.
\]
The first term vanishes as $r\to 0$, and the integral has a limit
$I\geq 0$ as $r\to 0$. If $I<\infty$ we are done. If $I=\infty$,
we apply l'H\^{o}spital's rule, and
\beas
0
&\leq&
\lim_{r\to 0} r^2 \int_r^1 \zeta\left(\frac{m(s)}{s^2}\right)\, ds
= \frac{1}{2}\lim_{r\to 0} r^3 \zeta\left(\frac{m(r)}{r^2}\right)
\leq \frac{1}{2}\lim_{r\to 0} r^3 \zeta\left(\frac{m(1)}{r^2}\right)\\
&\leq& 
\lim_{r\to 0} r^3 \frac{m(1)}{r^2} =0
\eeas
as desired.
We can assume that the steady state is non-trivial so that
$m(r)>m_0>0$ for $r$ large, and hence $\Psi_L$ is 
strictly increasing for $r$ large.
Together with its behavior at $r=0$ this implies 
that $\Psi_L$ attains a global minimum at some $r_L >0$.
We show that $\Psi_L'$ has at most one zero so that $r_L$ is unique
and $\Psi_L$ is strictly decreasing on $]0,r_L[$ and strictly increasing on
$]r_L,\infty[$. Altogether, these properties of $\Psi_L$ imply that each level
curve for an energy value above $\Psi_L(r_L)$ is a smooth connected curve
which does not contain the unique stationary point 
$(r_L,0)$ of (\ref{charseff}). Each such level curve therefore consists
of a single orbit of (\ref{charseff}) as desired.

It remains to show that  $\Psi_L'$ has indeed at most one zero. 
By (\ref{Mradpoisson}),
$\Psi_L'(r) = 0$ is equivalent to the equation
\be \label{zerocond}
r^3 \zeta \left(\frac{m(r)}{r^2}\right) = L.
\ee
Since $L>0$ and $\zeta(0)=0$ no solutions exist with $m(r)=0$.
Since $m$ is increasing and non-trivial,  $m(r)>0$ on
some interval $]r_0,\infty[$, and in order to see that 
there exists at most one solution of (\ref{zerocond})
we show that the derivative of the left hand side is positive 
on $]r_0,\infty[$. Clearly,
\beas
\frac{d}{dr} \left[r^3 \zeta \left(\frac{m(r)}{r^2}\right)\right]
&=&
3 r^2 \zeta \left(\frac{m(r)}{r^2}\right) + 
r^3 \zeta' \left(\frac{m(r)}{r^2}\right)
\left(-2 \frac{m(r)}{r^3}+ 4 \pi \rho(r)\right)\\
&\geq&
3 r^2 \left(\zeta \left(\frac{m(r)}{r^2}\right)
-\frac{2}{3}\frac{m(r)}{r^2}\zeta' \left(\frac{m(r)}{r^2}\right)\right).
\eeas
We recall that $\zeta^{-1}(\tau) = \tau \mu(\tau)$ and that 
$\mu$ is increasing. Hence
\[
\tau (\zeta^{-1})'(\tau) = \tau\left(\mu(\tau) + \tau \mu'(\tau)\right) 
\geq \tau \mu(\tau) > \frac{2}{3} \zeta^{-1} (\tau),\ 
\tau >0.
\]
In terms of $\zeta$ this is equivalent to the estimate
\[
\zeta(\sigma) - \frac{2}{3}\sigma \zeta'(\sigma) >0,\ \sigma>0.
\]
We substitute $\sigma = m(r)/r^2$ and conclude that
\[
\frac{d}{dr} \left[r^3 \zeta \left(\frac{m(r)}{r^2}\right)\right] >0,\ r>r_0,
\]
as desired.  The proof is complete. \prfe

In order to avoid mostly technical complications we restrict
ourselves to the more specific ansatz
\be \label{trueansatz}
f(x,v)= \Phi(E_0 - E) L^l
\ee
Here $l>-1/2$ and $E_0$ is a cut-off energy above which
the distribution is to vanish, i.e., 
$\Phi$ has to vanish for negative arguments. 
Such a cut-off energy is necessary in order to obtain steady 
states with compact support.
As we will see in the last section,
a cut-off energy is also necessary in order to obtain finite
mass for the case $\alpha <1$, but not in the genuine MOND case $\alpha =1$.
We make the following technical
assumptions on $\Phi$.

\smallskip

\noindent
{\bf Assumptions on $\Phi$.} 
$\Phi:\R \to [0,\infty[$ is measurable,
$\Phi (\eta) = 0$ for $\eta < 0$, and 
$\Phi > 0$ a.\ e.\ on some interval $[0,\eta_1]$ with $\eta_1>0$.
Moreover, there exists $\kappa > -1$ such that
for every compact
set $K\subset \R$ there exists a constant $C > 0$ such that
\[
\Phi(\eta) \leq C \eta^\kappa,\ \eta \in K.
\]
If we substitute the ansatz (\ref{trueansatz}) into
(\ref{rhodefss}) then in terms of $y:= E_0-U$,
\be \label{rhoyrel}
\rho(r) = r^{2 l} g(y(r))
\ee
where
\be \label{vpgdef}
g(y):= \left\{\begin{array}{ccl}
c_l \int_0^y \Phi(\eta)\, (y -\eta)^{l+1/2} d\eta&,&y>0,\\
0 &,& y\leq 0,
\end{array}\right.
\ee 
and $c_l>0$ is a constant.
The assumptions on $\Phi$ and
Lebesgue's dominated convergence theorem imply that
$g\in C(\R)\cap C^1(]0,\infty[)$. 
In terms of $y$ we have to solve the equation
\be\label{yeq}
y'(r) = - \zeta\left(\frac{m(r)}{r^2}\right)
\ee
where 
\be \label{mydef}
m(r) = m(r,y) = 4\pi \int_0^r s^{2l+2} g(y(s))\, ds;
\ee
notice that the right hand side of (\ref{yeq}) depends
on $y$ non-locally.
A solution launched by a central value
$y(0)=\open{y}\,>0$ gives a non-trivial steady state 
which has
finite mass and compact support if $y$, which is decreasing, has a zero.
In order to define the potential $U$ we have to make a
choice for the cut-off energy $E_0$. One possibility
is to let $E_0=0$ in which case 
$U = -y$ will have some unspecified limit at infinity.
If $y_\infty:=\lim_{r\to \infty} y(r) > -\infty$  which is the case 
if $y$ has a zero and $\alpha < 1$, but not for the genuine MOND
case, then we can define
$E_0 := y_\infty$ and $U:= E_0 -y$ to recover the standard
boundary condition (\ref{bc}) at infinity. We now show that 
(\ref{yeq}) does have a unique solution for every prescribed value of
$y(0)=\open{y}\,>0$. The question when this solution has a zero
so that the induced steady state has finite mass and compact support
is considered in the next section.
\begin{proposition}\label{existence}
Let $\open{y}\,>0$. Then (\ref{yeq}) has a unique
solution $y\in C^1([0,\infty[)$ with 
$y(0)=\open{y}$. This solution is strictly decreasing,
$y'(0)=0$, and $y\in C^2(]0,\infty[)$.
\end{proposition}
{\bf Proof.}
For $\delta > 0$ we define the set of functions
\[
Y := \{ y\in C([0,\delta]) \mid 
\open{y}\,/2\leq y(r) \leq \open{y}\,,\ r\in [0,\delta]\},
\]
a bounded, closed, convex subset of the Banach space $C([0,\delta])$ equipped
with the sup norm. The operator
\[
T(y)(r):= \open{y} - \int_0^r \zeta\left(\frac{m(s,y)}{s^2}\right)\, ds
\]
is defined on $Y$ and maps $Y$ into itself, provided $\delta$
is sufficiently small; note that for $y\in Y$ and $r\in [0,\delta]$,
\be \label{moverrsqest}
\frac{4 \pi g(\open{y}\,)}{2l+3} r^{2l+1} \geq
\frac{m(r,y)}{r^2}=\frac{4\pi}{r^2}\int_0^r s^{2+2l} g(y(s))\, ds \geq 
\frac{4 \pi g(\open{y}\,/2)}{2l+3} r^{2l+1}
\ee 
and recall that $2l+1>0$.
Moreover, $T$ is continuous,
and $T(Y)$ is a bounded and equicontinuous subset of $C([0,\delta])$.
By the Arzela-Ascoli Theorem and Schauder's Fixpoint Theorem
there exists a fixed point $y\in Y$ of $T$ which is
differentiable and satisfies (\ref{yeq}) on $[0,\delta]$ together
with the desired initial condition. 
The function $\zeta$ is in general not Lipschitz near $0$ so that
the mapping $Y$ need not be a contraction, but the solution is unique
on $[0,\delta]$, provided $\delta>0$ is sufficiently small.
To see this we first observe that for $\sigma\in [0,1]$,
\[
0\leq \frac{d}{d\sigma} \zeta^2(\sigma) 
=
\frac{2 \zeta(\sigma)}{\mu(\zeta(\sigma))+\zeta(\sigma)\mu'(\zeta(\sigma))}
\leq 
\frac{2 \zeta(\sigma)}{\mu(\zeta(\sigma))}\leq C\zeta(\sigma)^{1-\alpha}
\leq C\zeta(1)^{1-\alpha}
\]
so that $\zeta^2$ is Lipschitz on $[0,1]$. 
Consider two solutions $y$ and $\tilde y$
of the above initial value problem. Then
\beas
|y'(r) -\tilde y'(r)|
&=&
\frac{\left|\zeta^2\left(\frac{m(r,y)}{r^2}\right)-
       \zeta^2\left(\frac{m(r,\tilde y)}{r^2}\right)\right|}
{\zeta\left(\frac{m(r,y)}{r^2}\right)+
       \zeta\left(\frac{m(r,\tilde y)}{r^2}\right)}\\
&\leq&
C \zeta\left(\frac{m(r,y)}{r^2}\right)^{-1}
\left|\frac{m(r,y)}{r^2} - \frac{m(r,\tilde y)}{r^2}\right|.
\eeas
If $\delta$ is sufficiently small, (\ref{moverrsqest})
and the asymptotic behavior
of $\zeta$ imply that
\[
\zeta\left(\frac{m(r,y)}{r^2}\right) \geq C r^{(2l+1)/(1+\alpha)}
\]
with a positive constant $C>0$ which does not depend on $y, \tilde y$, 
$\delta$, or $r$.
Since $g$ is Lipschitz on $[\open{y}\,/2,\open{y}\,]$,
\[
\left|\frac{m(r,y)}{r^2} - \frac{m(r,\tilde y)}{r^2}\right|
\leq
C r^{2l+1} \max_{0\leq s\leq r}|y(s) - \tilde y(s)|
\]
so that altogether,
\[
|y(r) - \tilde y(r)|
\leq
\int_0^r|y'(s) -\tilde y'(s)|ds
\leq
C \delta^{1+\alpha (2l+1)/(\alpha+1)}
\max_{0\leq s\leq \delta}|y(s) - \tilde y(s)|
\]
where the constant $C>0$ does not depend on $y, \tilde y$, or $\delta$.
It follows that $y=\tilde y$
on $[0,\delta]$, provided $\delta$ is sufficiently small.
 
In order to extend this unique local solution
we observe that
in terms of the dependent variables $y$ and $m$ the equation (\ref{yeq})
is recast into the non-autonomous
first order system of ordinary differential equations
\[
y' = - \zeta\left(\frac{m}{r^2}\right),\ m' = 4 \pi r^{2+2l} g(y). 
\]
Prescribing positive data for $y$ and $m$ at some positive radius $r>0$
yields a unique local $C^1$ solution. The maximally extended solution
is strictly decreasing. Either it is bounded from below by zero
or it has a zero to the right of which $g(y)=0$ and $m=const$,
and in both cases the solution extends to $[0,\infty[$.
The right hand side of (\ref{yeq}) is continuously differentiable
for $r>0$, and it converges to zero for $r\to 0$. The proof is complete.
\prfe

\noindent
Given a solution of (\ref{yeq}) as obtained in Proposition~\ref{existence},
\[
f(x,v) = \Phi\left(y(r) - \frac{1}{2} |v|^2\right) |x\times v|^{2l}
\]
defines a spherically symmetric
steady state of (\ref{vlasov}), (\ref{rhodef}), (\ref{Mpoisson}). 
 
\section{Compact support and finite mass}
\label{sec_finite}
\setcounter{equation}{0}

\begin{theorem} \label{finiteradmass}
Let $f$ be a non-trivial, spherically symmetric steady state of the
MOND\-ian Vlasov-Poisson system
(\ref{vlasov}), (\ref{rhodef}), (\ref{Mpoisson}) as obtained in the
previous section.
\begin{itemize}
\item[(a)]
In the genuine MOND case $\alpha = 1$ the steady state
has finite mass and compact support.
\item[(b)]
In the general case the steady state
has finite mass and compact support, provided that $\Phi$ 
satisfies the additional assumption
\[
\Phi(\eta) \geq c \eta^k \ \mbox{for}\ \eta \in ]0,\eta_0[
\]
with parameters $c>0$, $\eta_0>0$, and $-1<k<l+3/2$.
\end{itemize}
\end{theorem}
{\bf Proof.} 
Since the steady state is non-trivial with
$g(y(0))>0$ and since the mass function $m$ is increasing
it follows that in the general case,
\[
m(r) \geq m(1)>0,\ r\geq 1,
\]
and since $\zeta$ is increasing,
\be \label{simpleest}
y'(r) \leq -\zeta\left(\frac{m(1)}{r^2}\right),\ r\geq 1.
\ee
In the genuine MOND case $\alpha =1$
the asymptotic behavior (\ref{zetaprop}) of $\zeta$ implies that
there exists some constant $\sigma_0 >0$ such that
\[
\zeta(\sigma) \geq \frac{1}{2} \sigma^{1/2},\ 0\leq \sigma \leq \sigma_0.
\]
We can choose $r_0 \geq 1$ such that 
\[
\frac{m(1)}{r^2}\leq \sigma_0,\ r\geq r_0,
\]
and hence
\[
y'(r) \leq -\frac{1}{2} \frac{m(1)^{1/2}}{r},\ r\geq r_0
\]
This estimate implies that $\lim_{r\to\infty} y(r) = -\infty$, in
particular $y(R)=0$ for some $R>0$ and $y(r) <0$
for $r>R$. By (\ref{rhoyrel}), $\rho(r)=0$ for $r>R$, 
and the proof of part (a) is complete.

Let us consider the general case;
in what follows $C>0$ denotes a constant which 
can change from line to line and which does not depend or $r$. 
Clearly, (\ref{zetaprop}) implies that 
$
\zeta(\sigma) \geq C \sigma,\ \sigma \geq 0,
$ 
and hence 
\[
y'(r) \leq - C \frac{m(r,y)}{r^2},\ r>0.
\]
We can now rely on the argument in \cite[Lemma~3.1]{RammRein} to conclude that
$y$ again has a zero. We include the short argument for the sake of 
completeness.
The monotonicity of $g$ and $y$ imply that
\[
y'(r) \geq -C r^{2l+1} g(y(r)).
\]
Hence for all $r>0$,
\be \label{key}
\int^{\open{y}}_{y(r)} \frac{d\eta}{g(\eta)} = 
- \int_0^r \frac{y'(s)}{g(y(s))}ds 
\geq  C \int_0^r s^{2l +1} ds 
= C r^{2l+2}.
\ee
We need to show that 
$y_\infty := \lim_{r\to\infty} y(r) < 0$; the limit exists by monotonicity.
If $y_\infty>0$ then (\ref{key}) gives a contradiction since the
left hand side is then bounded by $\open{y}\,/ g(y_\infty)$. 
It remains to
derive a contradiction from the assumption that $y_\infty =0$. 
The assumption in (b) and (\ref{vpgdef})
imply that
\be \label{gest}
g(y)
\geq C y^{n+l},\ 0\leq y \leq \eta_0,
\ee
where $0<n=k+3/2 < 3+l$ by the assumption on $k$. By 
the assumption that $y_\infty =0$,
$0<y(r)< \eta_0$ for $r$ sufficiently large. 
We split the left hand side
in (\ref{key}) accordingly, and by (\ref{gest}),
\[
C r^{2l+2} 
\leq \int^{\eta_0}_{y(r)}\frac{d\eta}{\eta^{n+l}} + 1.
\]
We compute the integral and multiply by $y(r)^{2l+2}$ to find that
\[
C \left(r y(r)\right)^{2l+2} \leq \
\left|(\eta_0)^{1-l-n}- y(r)^{1-l-n}\right|\,y(r)^{2l+2} 
+ y(r)^{2l+2};
\]
assume for the moment that $n+l\neq 1$. 
Since $n<l+3$, the right hand side of this estimate goes to zero
as $r\to \infty$. 
The same is true if  $n+l=1$, since then the integral yields
$\ln y(r)$ and $2l+2>0$. But (\ref{simpleest}),
the fact that $\zeta(\sigma) \geq C \sigma$,
and the assumption $y_\infty =0$ imply that the left hand 
side is bounded from below by
a positive constant. This contradiction
completes the proof.
\prfe

\noindent
{\bf Remark.} In the Newtonian case it is known that
an ansatz of the form
\be \label{polytr}
f(x,v) = (E_0 -E)_+^k L^l;
\ee
where the subscript $+$ denotes the positive part
leads to finite mass and compact support,
provided $k, l > -1$ with
$k+l+3/2 \geq 0$ and $k < 3 l + 7/2$ while the corresponding steady states
have infinite mass for $k > 3 l + 7/2$.
For $l+3/2<k<3 l + 7/2$ these polytropic states have finite mass and radius,
but are structurally unstable in the sense of \cite[p.~378]{RR00}.
The robust finite radius proof based on \cite[Lemma~3.1]{RammRein}
breaks down for $k>l+3/2$, and it is questionable
whether part (b) of the theorem persists for  polytropes with
$k>l+3/2$.

\section{Additional results}
\label{sec_final} 
\setcounter{equation}{0}
\subsection{Asymptotic behavior and potential energy}
We define the energy of the gravitational field with
potential $U$ by
\be \label{sdef}
S(U) := 
\frac{1}{2} a_0^2\int F(|\nabla U|^2/a_0^2)\,dx,
\ee
where we have restored the threshold parameter $a_0$ for easier comparison 
with (\ref{Mpoisson}), and
\[
F(\tau) := \int_0^\tau \mu(\sqrt{s})\, ds,\ \tau \geq 0.
\]
Formally, the modified Poisson equation (\ref{Mpoisson})
is the Euler-Lagrange equation of $S(U) + 4\pi \int U \rho$,
and $S(U)$ is part of the formally conserved total
energy of the MONDian Vlasov-Poisson system.
Hence it is desirable that $S(U)$ be defined
for steady states with compact support and finite mass,
and this issue is related to the asymptotic behavior
of the potential at infinity and the possibility
of restoring the standard boundary condition (\ref{bc}).
\begin{proposition} \label{limits}
Consider a non-trivial, spherically symmetric steady state of the
MONDian Vlasov-Poisson system.
\begin{itemize}
\item[(a)]
If $\alpha < 1$ and the steady state has finite mass, then
$\lim_{r\to \infty} U(r) < \infty$ and this limit can be taken
to vanish so that (\ref{bc}) holds. The integral $S(U)$ converges.
\item[(b)]
In the genuine MOND case $\alpha=1$, $\lim_{r\to \infty} U(r)=\infty$
and $S(U)=\infty$ even if the steady state has finite mass and compact support.
\end{itemize}
\end{proposition}
{\bf Proof.} Since the steady state has positive, finite mass,
\[
\lim_{r\to \infty} \frac{m(r)}{r^2} =0
\]
and by (\ref{zetaprop}) there exist constants $C_1, C_2 >0$ such that
\be \label{asest}
C_1 r^{-2/(1+\alpha)} \leq \zeta\left(\frac{m(r)}{r^2}\right)
\leq C_2 r^{-2/(1+\alpha)}
\ee
for $r$ sufficiently large.
Now we observe that the assumption (\ref{muass})
on $\mu$ implies that $F\in C^1([0,\infty[)$ and
\[
\lim_{\tau \to 0} F(\tau^2)\tau^{-2-\alpha} =\frac{2}{2+\alpha}.
\]
Hence $S(U)$ converges if and only if the function
$r^{2-2 (2+\alpha)/(1+\alpha)}=r^{-2/(1+\alpha)}$ is integrable
on some interval $[R,\infty[$ with $R>0$. This is the
case if and only if $\alpha \in [0,1[$ which proves the assertions
on $S(U)$. Moreover, (\ref{asest}) and (\ref{Mradpoisson}) imply that
\[
\lim_{r\to\infty} U(r) = U(R) + \int_R^\infty U'(s)\, ds
\]
is finite for $\alpha <1$ and infinite for $\alpha =1$.
The proof is complete. \prfe

\noindent
{\bf Remark.} The above proposition shows that the potential
$U$ is always confining in the genuine MOND case $\alpha=1$.
This is consistent with the fact that in that case
no restriction on
the ansatz (\ref{trueansatz}) is needed  to
guarantee finite mass and compact support of the resulting
steady state. We illustrate this fact further by
considering Maxwellian
distributions.
\subsection{Maxwellians}
We assume that the distribution function $f$ of a steady state is 
of the form
\be \label{maxwellf}
f(x,v) = e^{-E} = e^{-\frac{1}{2} |v|^2 - U(x)}.
\ee
Then the relation between the mass density and the potential becomes
\[
\rho(x)=(2\pi)^{3/2} e^{-U(x)},
\]
in particular, this relation has the regularity and monotonicity properties
required for Proposition~\ref{existence} so that the ansatz (\ref{maxwellf})
leads to corresponding steady states of the MONDian Vlasov-Poisson system.
It turns out that in the genuine MOND case such a Maxwellian steady state
has finite mass.
\begin{proposition} \label{maxwellfinite}
In the genuine MOND case $\alpha = 1$
the ansatz (\ref{maxwellf}) leads to spherically symmetric steady states
of the system (\ref{vlasov}), (\ref{rhodef}), (\ref{Mpoisson})
with finite mass.
\end{proposition}
{\bf Proof.} 
We recall that there exists some constant $\sigma_0 >0$ such that
\[
\zeta(\sigma) \geq \frac{1}{2} \sigma^{1/2},\ 0\leq \sigma \leq \sigma_0.
\]
Let us assume that a steady state of Maxwellian type has infinite mass.
Then there exists some $r_0 > 0$ such that $m(r_0) > 36$. We can choose
$R>r_0$ such that 
\[
\frac{m(r_0)}{r^2} < \sigma_0,\ r\geq R,
\]
and hence by the monotonicity of $m$ and $\zeta$,
\[
U'(r) = \zeta\left(\frac{m(r)}{r^2}\right) 
\geq \zeta\left(\frac{m(r_0)}{r^2}\right)
\geq \frac{m(r_0)^{1/2}}{2 r} =: \frac{c_0}{r},\ r\geq R,
\]
where $c_0 >3$. Integrating this implies that
\[
U(r) \geq U(R) - c_0 \ln R + c_0 \ln r,\ r\geq R,
\]
and therefore
\beas
m(r) 
&=& 
m(R) + (2\pi)^{3/2} 4 \pi \int_R^\infty s^2 e^{-U(s)}ds\\
&\leq& 
m(R)+ (2\pi)^{3/2} 4 \pi e^{c_0 \ln R - U(R)} \int_R^\infty s^{2-c_0} ds,\ r\geq R.
\eeas
Since $c_0 > 3$, the latter integral converges, and this contradicts 
the assumption that the total mass of the steady state is infinite.
\prfe

If $\alpha < 1$ then an ansatz of the form (\ref{maxwellf}) necessarily
leads to steady states with infinite mass. This is a corollary to
the following more general result which in turn is just an extension
of the corresponding Newtonian result \cite[Theorem~2.1~(a)]{RR00} to 
the present modified situation; for the sake of completeness we include 
the corresponding argument.
\begin{proposition} \label{cutoff}
Let $\alpha < 1$ and assume that an ansatz of the general form (\ref{ansatz})
leads to a spherically symmetric
steady state with finite mass. Then there exists a cut-off energy
$E_0$ such that $\Phi(E,L)=0$ for almost all $(E,L)$ with $E>E_0$.
\end{proposition}
{\bf Proof.} As shown in 
Proposition~\ref{limits}, $U$ is increasing with a finite 
limit $U_\infty$ at infinity. A simple change of variables shows that the total
mass $M$ of the steady state is given by
\beas
M
&=&
8 \pi^2 \int_0^\infty \int_{U(r)}^\infty \int_0^{2 r^2(E-U(r))}
\Phi(E,L) \frac{dL\,dE\,dr}{\sqrt{2(E-U(r)-L/2r^2)}}\\
&\geq&
8 \pi^2  \int_{U_\infty}^\infty  \int_0^\infty \Phi(E,L)
\int_{\sqrt{L/(2(E-U_\infty))}}^\infty \frac{dr}{\sqrt{2(E-U(r))}}
dL\,dE .
\eeas
For $E>U_\infty$ and $L>0$ the integral with respect to $r$ 
in the latter expression is infinite,
and hence $\Phi$ must vanish for such arguments. \prfe

\subsection{The MONDian Euler-Poisson system}
The results which we have discussed to far have
counterparts if matter is described as an ideal, compressible
fluid instead of a collisionless gas.
We replace the Vlasov equation
by the compressible
Euler equations and from the start restrict ourselves to the
spherically symmetric, time-independent case.
The pressure $p=p(r)$ is given in terms of the mass density
$\rho=\rho(r)$ via an equation of state
\be \label{eqofst}
p=P(\rho),
\ee
the velocity field vanishes,
and the static, spherically symmetric Euler equation reads
\be \label{eulerrad}
P'(\rho)\, \rho' + \rho\, U' = 0.
\ee
Supplemented with (\ref{Mradpoisson}) these equations constitute 
the stationary, spherically symmetric case of
the MONDian Euler-Poisson system; we refer to \cite{RammRein}
for its Newtonian analogue. 
We make the following technical assumptions on the equation of state.

\smallskip

\noindent
{\bf Assumptions on $P$.}
Let $P\in C^1([0,\infty[)$ be such that $P'>0$ on $]0,\infty[$,
\[
\int_0^1 \frac{P'(s)}{s}ds < \infty,\ \mbox{and}\
\int_0^\infty \frac{P'(s)}{s}ds = \infty.
\]
We define
\[
Q(\rho):= \int_0^\rho \frac{P'(s)}{s}ds,\ \rho\geq 0.
\]
Then $Q:[0,\infty[ \to [0,\infty[$ is one-to-one and onto,
$Q\in C([0,\infty[)\cap C^1(]0,\infty[)$ with $Q(0)=0$
and $Q'(\rho)= P'(\rho)/\rho$ for $\rho>0$. The Euler equation 
(\ref{eulerrad}) holds provided
\be \label{rhoUrel}
Q(\rho(r)) = c - U(r),\ r\geq 0,
\ee
with some integration constant $c$.
If we define
\be \label{epgdef}
g(y):=\left\{
\begin{array}{ccl}
Q^{-1}(y) &,& y>0,\\
0&,&y\leq 0,
\end{array} \right.
\ee
then $g\in C(\R)\cap C^1(]0,\infty[)$,
and writing $y=c-U$ we invert the relation (\ref{rhoUrel})
to read as in (\ref{rhoyrel}) with $l=0$ there.
Hence the static Euler-Poisson system is reduced to the same equation
(\ref{yeq}) with mass function defined by (\ref{mydef}) with $l=0$
and with $g$ defined by (\ref{epgdef}) instead of (\ref{vpgdef}).
We are therefore back to the same situation as for the MONDian
Vlasov-Poisson system. 
As in Proposition~\ref{existence}
we obtain for any choice of $y(0)=\open{y} >0$ a unique
solution of (\ref{yeq}). It induces a stationary, spherically 
symmetric solution of the MONDian Euler-Poisson system.
In the genuine MOND case $\alpha =1$ this steady state
always has compact support and finite mass, while for the
general case $\alpha \in [0,1]$ these properties hold provided
\[
P'(\rho)\leq c \rho^{1/n}
\]
for $\rho>0$ and small and with $0<n<3$. These assertions follow
exactly as in Theorem~\ref{finiteradmass}; notice that under the above
assumption
$Q(\rho) \leq C \rho^{1/n}$ for small values of $\rho$ 
which in turn implies that
$g$ defined in (\ref{epgdef}) satisfies the estimate (\ref{gest}) with $l=0$, 
and the proof can proceed as for the Vlasov case.

\end{document}